# Amplitude of Random Telegraph Noise in Junctionless FinFET with Different Channel Shape

Atabek E. Atamuratov,[a, †]   Mahkam M. Khalilloev,[a]   Ahmed Yusupov,[b]   Jean Chamberlain Chedjou,[c]
Kyandoghere Kyamakya [c]

[a] *Physics department, Urgench State University, Kh.Olimjan str.,14, Urgench, 220100, Uzbekistan*
[b] *Department of Electronics and Radiotechnics, Tashkent University of Information Technologies, A.Temur str.108 Tashkent, 200100, Uzbekistan*
[c] *Institut für Intelligente Systemtechnologien, University of Klagenfurt, Klagenfurt, 9020, Austria*
[†] *Corresponding author:* atabek.atamuratov@urdu.uz

The influence of the channel shape in a junctionless silicon-on-insulator finned field-effect transistor (JL SOI FinFET) on the amplitude of random telegraph noise (RTN) induced by single interface trapped charge has been simulated for the transistors with rectangular, trapezoidal, and triangular fin cross sections. The simulation of the RTN amplitude distribution along the channel induced by a single charge trapped at interface defect located at the fin top and at sidewall of JL SOI FinFETs with channels of different shapes is considered. It is established that at trapping the single charge at sidewall surface of the channel, the lowest RTN amplitude is seen for the triangular cross-section and the highest for the rectangular and trapezoidal cross-sections. At the single charge trapping at the top surface of the channel, the RTN amplitude is higher for the rectangle than for the trapezoidal cross-section.

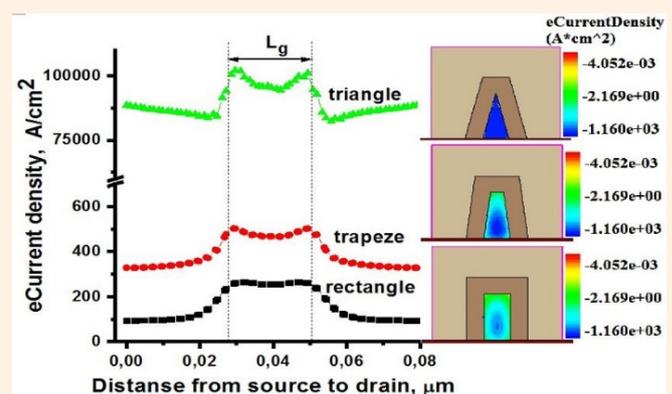

## I. INTRODUCTION

The scaling of metal-oxide-semiconductor field effect transistors (MOSFETs), especially junctionless (JL) finned field-effect transistors (FinFETs) [1], can cause various degradation effects. The short channel effects and the increase in the sensitivity of the drain current to charge trapping in gate oxide or at the interface are some of them. While the short channel effects in the JL-FinFET were thoroughly investigated [2–4], issues related to the influence of single oxide- or interface-trapped charges on both the drain current and induced random telegraph noise (RTN) signals in transistors [5] are studied to a much lesser extent. Some previous studies have focused on the dependence of RTN amplitude on the transistor manufacturing technology, temperature, gate and drain voltages in accumulation mode [6, 7], and the dependence of JL FinFET operating efficiency in ultra-low power logic on the level of channel doping, the dielectric constant of the spacer and the channel resistance [8]. To the best of our knowledge, very few papers have reported on the influence of technological factors such as variations in geometric parameters and deviations from preset shapes of nano-sized JL FinFETs on their noise properties. However, these variations can lead to significant deviations in transistor hybrid parameters and characteristics. In particular, it was found that variations in the characteristics, parameters, and short channel effects depend both on the channel (fin) shape and on the dimensions of the FinFETs operating in inversion mode [9–11]. Besides it is shown that, variations of the channel shape can considerably influence on RTN amplitude in inversion mode FinFET [12]. Therefore, to compare the influence of different channel shapes to RTN amplitude, it is interesting to consider RTN in JL FinFET, operating in accumulation mode, with different channel shape.

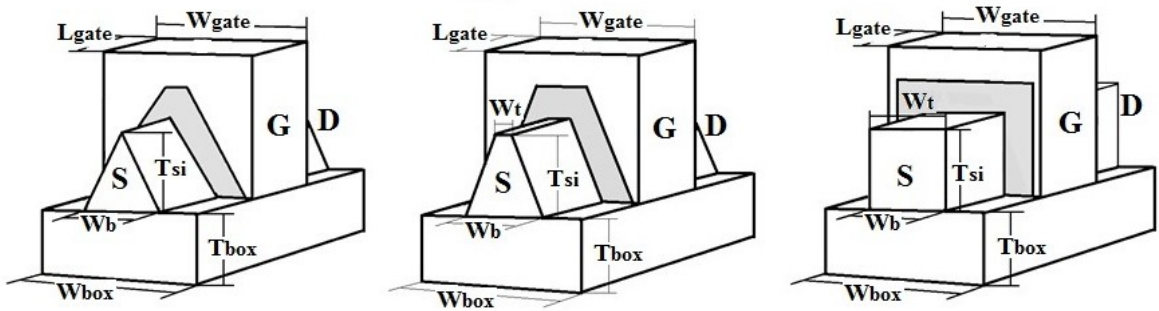

Figure 1: The structures of simulated junctinoless FinFETs with different channel shapes.

## II. SIMULATION METHODOLOGY

Transistors with triangular, trapezoidal, and rectangular cross-sections of the channel were considered. The influence of traps located at the top was considered only for transistors with trapezoidal and rectangular cross-sections. Various simulations were carried out for a specific size of the gate (G) length ($L_{gate}$ = 22 nm) and different channel shapes (see Figure 1). The gate oxide used was a hafnium oxide ($HfO_2$) layer with an equivalent thickness of $t_{eff}$ = 1.2 nm. The buried oxide layer had a width $W_{box}$ = 48 nm and a thickness $T_{box}$ = 145 nm. The fin shape was characterized by the top width $W_t$ of dimension 5 nm for the trapezoidal cross-section. The fin with a base of width $W_b$ = 10 nm was doped with boron of concentration $5 \times 10^{18}$ cm$^{-3}$.

Using the Advanced TCAD Sentaurus program package, various three-dimensional simulations were carried out in the frame of a drift-diffusion model. The physical model took into account the dependence of the carrier mobility on the doping level, the saturation of the carrier velocity and the influence of the normal field component on the drain current.

Since the transistor was nanoscale in size, it was also necessary to consider the quantum confinement effects. In the present work, quantum corrections were used with respect to the density gradient, which were preferred in the context of drift-diffusion simulations. The adopted model was calibrated according to the experimental data of Barraud et al. [13] and results of calibration is presented in our previous work [14].

## III. SIMULATION RESULTS AND DISCUSSION

The simulation results show that the RTN amplitude depends on the transistor channel shape. The underlined dependence can be observed in the sub-threshold range of the gate voltages in Figure 2. As clearly depicted in Figure 2, the RTN amplitude for the triangular cross-section is approximately five times higher than for the trapezoidal and rectangular cross-sections when the single charge is trapped in the center at sidewall and at the gate voltage overload of −0.4 V. The RTN amplitude distribution along the channel during the single charge trapping at side wall as well as at the top surface of the channel was considered (or determined) for both the gate threshold voltage and a gate voltage overdrive of −0.4 V (see Figure 3) for all shapes of the cross-sections at stake. In all depicted distributions, the maximum and minimum RTN amplitudes can be seen in rectangular and triangular cross-sections, respectively. This result is correlated with a more noticeable influence of the single trapped charge at low drain current than at high current. The RTN amplitude is the relative change of the current in the channel $\Delta I/I$. With a high current density, the absolute change of the current ($\Delta I$) under influence of the same charge is smaller than with a low current density. Obviously, this is associated with a more high screening of the influence of the trapped charge at high current densities. Therefore, the RTN amplitude is lower at high current densities. In the case of a triangular cross-section, the RTN amplitude is lowest due to the highest current density in the channel.

The width at the bottom and the height of the channel are the same for all cross-sections considered and consequently the cross-sectional area of the triangle is the smallest and, therefore, the current density is the highest. The channel current density is also influenced by the gate field over the channel, although the cross-sectional area is critical. In case of the rectangular and trapezoidal cross-sections, the gate field side wall components mainly compensate one another. The vertical component of the gate field causes a nonuniform distribution of the current along the height and shifts the center of the area with the maximum current density to the lower part of the channel (Figure 4, inset). With a triangular cross-section of the channel, the vertical component of the gate field is missing, which leads to a more or less uniform distribution of the current along the height (Figure 4, inset).

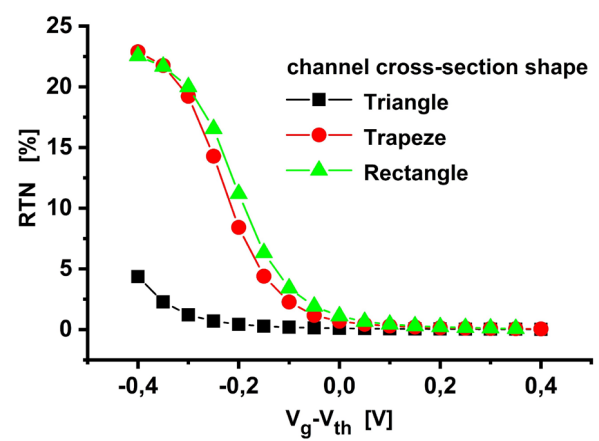

Figure 2: RTN amplitude dependence on the gate voltage overdrive for JL SOI FinFETs with different channel shapes. RTN is induced by single charge trapped at the center of side wall surface of the transistors.

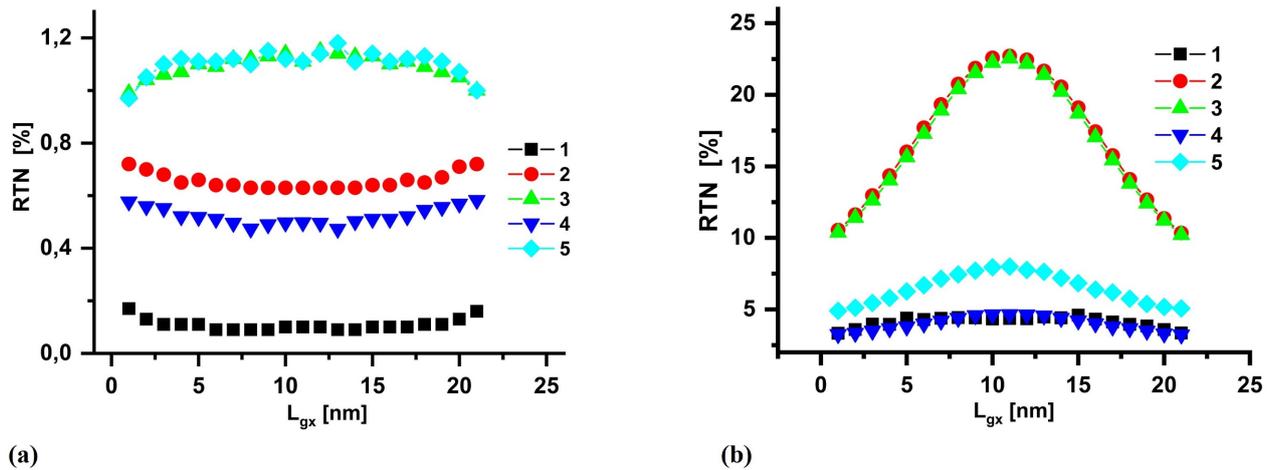

Figure 3: RTN amplitude distributions along the channel induced by single charge trapped at the side wall surface (1, 2, 3) as well as at the top surface (4, 5) for different channel shapes: triangle (1), trapeze (2, 4), and rectangle (3, 5) at the gate voltage overdrive of 0 V (a) and −0.4 V (b).

Besides, the influence to the RTN amplitude is greater for short distances between the trapped charge and the area of the maximum current density in the channel than for long distances. As it is clearly depicted from Figure 4, the drain current density along the entire channel is highest for triangular cross-sections and lowest for rectangular cross-sections.

As mentioned above, the gate field affects the current distribution across the channel. The current density increases in the direction from the surface to the center of the channel. An area with the maximum current density can be seen on the rectangular and trapezoidal cross-section of the channel in a defined position of the channel cross section (area A in Figure 5). The level of impact of the trapped single charge on the current in the channel depends on the distance $d$ between the trapped charge and the center of area A. In the case of trapping the single charge in the center of the side wall of the trapezoidal cross-section channel, the distance $d$ is the same as in the case of the rectangular cross-section channel ($d2 \approx d4$, Figure 5). Therefore, the RTN amplitude for these cross-sections is roughly the same, but higher than in the case of the triangular cross-section (Figure 2).

In the carrier density distribution under the gate along the channel there is a minimum in the middle of the channel for all shapes of cross-sections. Consequently, a maximum of the RTN amplitude is created in the middle of the channel. As it appears from Figure 3, the underlined maximum is very noticeable at a gate overload of −0.4 V. The RTN amplitude distribution along the channel at the top surface was considered only for trapezoidal and rectangular cross-sections of the channel (see the plots in Figure 3). It can be seen (from Figure 3) that the single charge trapped at the upper/top surface induces RTN with a lower amplitude than the single charge trapped at side wall of the channel for both cross-sectional shapes. This result is correlated with long distances from the area of maximum current density (in the channel) to the single charge trapped at the top surface than with the charge trapped at side wall surface of the channel (Figure 5).

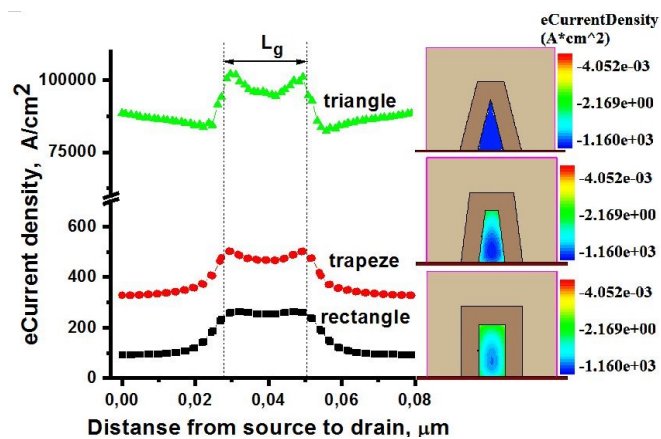

Figure 4: Distribution of the electron current density along the fin length at the middle of the fin at the gate voltage overdrive of −0.4 V and $V_{ds}$ = 10 mV for different channel shapes. In the inset, it is shown the distribution of the electron current density at the cross section of the channel center for different channel shapes.

The distance between the area of maximum current density in the channel and the single charge trapped at the top surface is shorter for a rectangular cross-section than for a trapezoidal one ($d1 < d3$). Besides, the maximum current density for the trapezoidal cross-section is higher. As a result, the RTN amplitude in the case considered above is higher for the rectangular cross-section than for the trapezoidal cross section (see the plots in Figure 3). The distances $d$2 and $d$4 are approximately equal and therefore the RTN amplitudes induced by the single charges trapped at the sidewall surfaces for both channel shapes are approximately equal (see the plots in Figure 3).

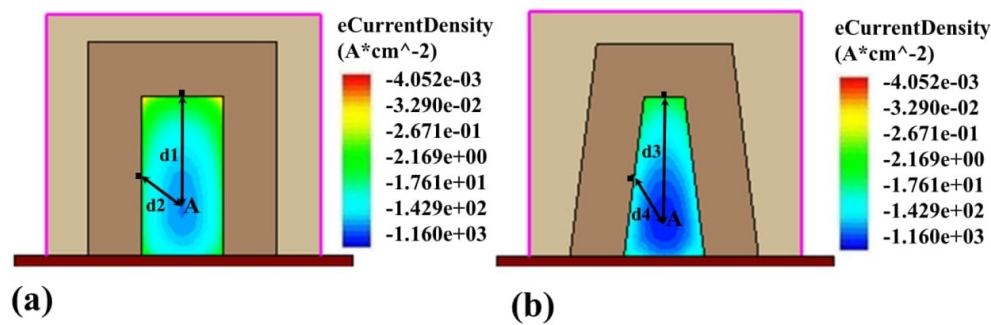

Figure 5: Distribution of electron current density at cross section of the channel center for rectangular (a) and trapezoidal (b) cross-sections at absent of trapped charge. $d$1 and $d$2 is distances between area with maximal current density in the channel cross section and prospective single charges trapped at the top and side wall surfaces, respectively, for rectangular cross section, $d$3 and $d$4 the same distances for trapezoidal cross-section.

## IV. CONCLUSION

Our study has revealed that the RTN amplitude in the junctionless SOI FinFET in the subthreshold mode is relatively high and depends on the channel shape and the position of the single trapped charge. At the single charge trapping at sidewall surface of the channel, the lowest RTN amplitude is seen/observed for the triangular cross-section and the highest for the rectangular and trapezoidal cross-sections. At the single charge trapping at the top surface of the channel, the RTN amplitude is higher for the rectangle than for the trapezoidal cross-section. This result is correlated with a shorter distance between the area/region of maximum current density in the channel and the trapped charge in the case of a rectangular cross-section. Thus, the variation of the channel shape can lead to a noticeable change in the RTN amplitude in the junctionless SOI FinFETs.

### Acknowledgments

This research was funded by Ministry of Innovative Development of the Republic of Uzbekistan, grant number OT-F2-67.

### References

[1] I. Ferain, C. A. Colinge, and J.-P. Colinge, Nature 479, 310 (2011).
[2] B. C. Paz, F. Ávila-Herrera, A. Cerdeira, and M. A. Pavanello, Semicond. Sci. Technol. 30, 055011 (2015).
[3] N. Jaiswal and A. Kranti, IEEE Trans. Electron Devices 65, 3669 (2018).
[4] P. Razavi, N. D. Akhavan, R. Yu, G. Fagas, I. Ferain, and J. P. Colinge, Extended Abstracts of the 2011 International Conference on Solid State Devices and Materials (Nagoya, 2011) p. 106.
[5] J. P. Campbell, L. C. Yu, K. P. Cheung, J. Qin, J. S. Suehle, A. Oates, and K. Sheng, 2009 IEEE International Conference on IC Design and Technology (Austin, 2009) p. 17.
[6] M.-L. Fan, S.-Y. Yang, V. P.-H. Hu, Y.-N. Chen, P. Su, and C.-T. Chuang, Microelectron. Reliab. 54, 698 (2014).
[7] A. N. Nazarov, I. Ferain, N. D. Akhavan, P. Razavi, R. Yu, and J. P. Colinge, Appl. Phys. Lett. 98, 092111 (2011).
[8] D. Roy and A. Biswas, Superlattices Microstruct. 97, 140 (2016).
[9] A. E. Abdikarimov, A. Yusupov, and A. E. Atamuratov, Tech. Phys. Lett. 44, 962 (2018).
[10] A. E. Atamuratov, A. Abdikarimov, M. Khalilloev, Z. A. Atamuratova, R. Rahmanov, A. GarciaLoureiro, and A. Yusupov, Nanosyst.: Phys. Chem. Math. 8, 71 (2017).
[11] A. Abdikarimov, G. Indalecio, E. Comesaña, A. J. Garcia-Loureiro, N. Seoane, K. Kalna, and A. E. Atamuratov, 2014 International Workshop on Computational Electronics (Paris, 2014) p. 1.
[12] A. E. Abdikarimov, Tech. Phys. Lett. 46, 494 (2020).
[13] S. Barraud, M. Berthome, R. Coquand, M. Casse, T. Ernst, M.-P. Samson, P. Perreau, K. K. Bourdelle, O. Faynot, and T. Poiroux, IEEE Electron Device Lett. 33, 1225 (2012).
[14] M. M. Khalilloev, B. O. Jabbarova, and A. A. Nasirov, Tech. Phys. Lett. 45, 1245 (2019).

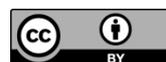